\documentclass[5p,twocolumn,times,number]{elsarticle}

\usepackage{graphicx}
\usepackage{amsmath}
\usepackage{epsfig}
\begin{document}
\begin{frontmatter}
\title{Test of a LYSO matrix with an electron beam
between 100 and 500 MeV for KLOE-2}
\author[add1]{M.~Cordelli}
\author[add1]{F.~Happacher}
\author[add1]{M.~Martini}
\author[add1]{S.~Miscetti\corref{cor}}
\ead{Stefano.Miscetti@lnf.infn.it}
\author[add1]{I.~Sarra}
\author[add2]{M.~Schioppa}
\author[add2]{S.~Stucci}
\author[add3]{G.~Xu}
\cortext[cor]{Corresponding author}
\address[add1]{Laboratori Nazionali di Frascati dell'INFN, Frascati, Italy}
\address[add2]{INFN and department of physics, University of Calabria, 
Cosenza,Italy}
\address[add3]{Institute of High Energy Physics of Academia Sinica, Beijing, China}

\begin{abstract}
The angular coverage extension of the KLOE-2 electromagnetic calorimeter,
from a polar angle of 20$^{\circ}$ down to $8^{\circ}$, will
increase the multiphoton detection capability of the experiment 
enhancing the search reach for rare kaon, $\eta$ and  $\eta'$ prompt 
decay channels. The basic layout of the calorimeter extension
consists of two small barrels of LYSO crystals
readout with APD photosensors aiming to achieve a timing
resolution between 300 and 500 ps for 20 MeV photons. 
The first test of a (5.5$\times$6$\times$13) cm$^3$ prototype for such
a detector  was carried out in april 2009 at the Beam Test Facility
of Laboratori Nazionali di Frascati of INFN with an electron
beam from 100 to 500 MeV. In the selected energy range, we 
measured a light yield of 500$\div$800 p.e./Mev,  an energy resolution 
which can be parametrized as $0.05 \oplus 0.01/({\rm E/GeV}) \oplus 0.015/\sqrt{\rm{E/GeV}}$, 
a position resolution of 2.8 mm and a timing resolution of 200$\div$300 ps.
\end{abstract}


\end{frontmatter}

\section{Introduction}
In the last years, a new machine scheme based on the Crab-waist and
a large Piwinsky angle has been proposed and tested~\cite{Dafne} to
improve the reachable luminosity at the Frascati $\phi$-factory, DA$\Phi$NE,
an e$^{+}$e$^{-}$ collider running at the center of mass energy of the $\phi$ 
resonance. The success of this test motivated the startup of a new experiment,
named KLOE-2~\cite{kloe2_loi}, which aims to complete the 
KLOE~\cite{kloe_cimento} physics program and perform a new set of 
interesting measurements. The first running phase
will start at the end of 2009 with the goal of collecting $\sim$ 5 fb$^{-1}$ 
in one year. For this first run, the modification of the KLOE
detector will be minimal with the only introduction of  a $\gamma \gamma$ 
tagging system. A second phase, for a longer data taking, 
will require another set of upgrades all concentrated around the beam-pipe 
consisting of: (i) an inner tracker, IT,  based on Cylindrical GEM 
technology, (ii) a tile calorimeter surrounding the inner 
quadrupoles , QCALT , 
and (iii) a calorimeter between the interaction point, IP, and the first
inner quadrupole. In the following, we describe the work done
to design and test the possibile solution for the latter detector.

\begin{figure}
\centering
\includegraphics[width=0.8\linewidth]{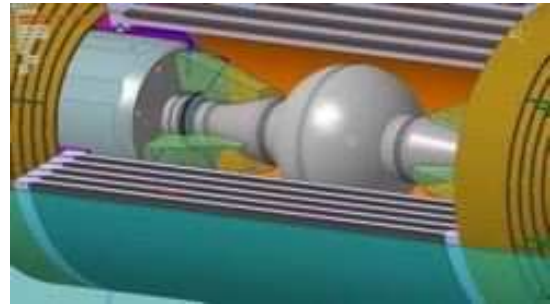}
\caption{Zoomed-view of the IP region. The area available for
the new calorimeter lies between the inner sphere and the closest
quadrupoles.}
\label{ccalt_region}
\end{figure}

In Fig.~\ref{ccalt_region}, we show a zoomed-view of the available
region around the IP which can be used to extend the angular
coverage of the main electromagnetic calorimeter, today limited
at a polar angle of $\sim$ 20$^{\circ}$, with the addition 
of a new dedicated calorimeter. Assuming to be able
to lower the minimum polar angle for photon detection
down to 8$^{\circ}$, this will enhance the multiphoton detection
capability of the detector for the search of rare decays
of kaons, $\eta$ and $\eta'$ mesons. As an example,
in 2007 the KLOE experiment has measured a 
BR($K_S \to \gamma \gamma$)~\cite{ksgg} in good agreement with
O(p$^4$) calculation of Chiral Perturbation Theory but 
which differs of three  standard deviations from the same
measurement carried out by the NA48 experiment.   
The proposed upgrade will allow to improve
the S/B ratio of a factor of three thus allowing to solve
this puzzle. Other searches or branching ratio determinations, 
such for instance $K_s \to 3 \pi^0$ or  $\eta \to \pi^0 \gamma \gamma$,  
will directly benefit from this extension.
The only available area to place a calorimeter lies between 
the end of the  spherical beam pipe, of 10 cm radius,  
and the first quadrupole, positioned at 30 cm distance from the IP.

\section{CCALT: a Crystal Calorimeter with Time}

\begin{figure}
\centering
\includegraphics[width=0.7\linewidth]{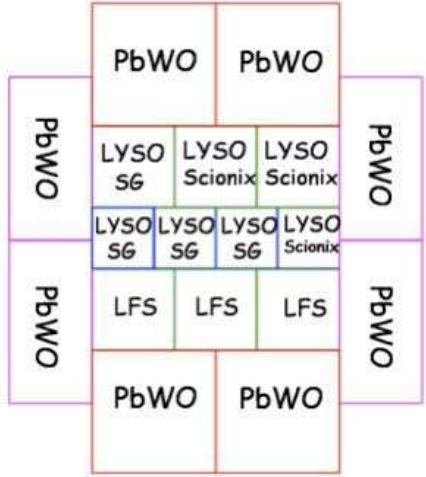}
\caption{Composition of crystals used for the matrix prototype.}
\label{matrix_scheme}
\end{figure}
The discussion of the previous section indicates that this
calorimeter has to be very dense, with a small value of
radiation length, $X_0$,  and Moliere radius, $R_m$,
not hygroscopic and with a large light output to
improve photon detection efficiency at low energy
(from 20 to 500 MeV). Moreover, the calorimeter has 
to be extreemely fast in order to allow prompt photon 
reconstruction in an environment with a large background 
rate ($\sim$~1$\div$ 5  MHz) of secondary showers generated 
by off-axis e$^{+}$, e$^{-}$ coming from
intra-bunch scattering ({\em Touschek effect}). 
Preliminary simulation studies indicates the need
to reach a time resolution of 300$\div$ 500 ps for 
20 MeV photons.

A suitable solution is offered by a crystal calorimeter with good timing 
performances, named CCALT. A first detector layout consists of 
two concentrical barrels of 24 crystals each,  with transversal 
dimensions of 2$\times$2 cm$^2$ and longitudinal length between 13 and 15 cm.
The best crystal choice matching the requirements is provided
by the new generation of Cerium doped Lutetium Yttrium Orthosilicate, 
LYSO, which has  $X_0$ and $R_M$ values
(1.1 and 2 cm) comparable to the ones (0.9 and 2 cm) 
of the Lead Tungstanate, PbWO$_{4}$, with the advantage of a 
much larger light yield ($\times$ 300). On the negative side, 
LYSO shows a scintillation emission time ($\tau_{LYSO} = 40$ ns)
slower than PbWO$_{4}$ ($\tau_{PbWO}= 10$ ns). However, from the basic 
scaling law of the timing  resolution, $\sigma_t = \tau/\sqrt{{\rm Np.e.}}$, 
we expect LYSO to be a factor four more performant than PbWO$_{4}$.

In the final location of the CCALT inside KLOE-2,
the presence of an axial magnetic field of 0.52 kGauss forces
the usage of silicon based  photodetectors. 
Due to the high photon yield, the readout with APDs
is a valid solution since, at the lowest photon energy of 20 MeV, 
the collected photoelectrons will be around $\sim$ 10.000  which 
corresponds to $\sim$ 12 pC assuming an average gain of 300 and an
amplification stage of $\times 25$,  well matching the ADC
sensitivity of the KLOE calorimeter (100 fC/count). 
In the following,  we specifically 
considered only the Hamamatsu S8664-55, which has an
active area of 0.5$\times$0.5 cm$^2$, fast timing
characteristics and a quantum efficiency between 65 and 85\%
in the wavelenght range of interest (390$\sim$ 500 nm) for
the LYSO emission spectra.

In march 2009, we have built a medium size crystal matrix prototype 
with transversal radius larger than 2 $R_m$,  longitudinal 
dimensions being constrained by budget limits to be between 
13 and 15 cm (corresponding to 11 $\div$ 12 $X_{0}$ of 
longitudinal containment).
The prototype consists of an inner matrix of 10 LYSO crystals
readout by APD and an outer matrix, for leakage recovery, 
composed by 8 PbWO$_4$ cristals readout by standard Hamamatsu 
Bialcali photomultipliers of 1,1/8'' diameter.
To test the quality of the crystals offered by different
vendors, the inner matrix has been assembled in three rows
(see  Fig.~\ref{matrix_scheme}) composed (from bottom to top)
as follows:
\begin{itemize}
\item 3 LFS crystals from Zecotek of 2$\times$2$\times$13 cm$^3$,
\item 2 LYSO St.Gobain crystals of  1.5$\times$1.5$\times$15 cm$^3$,
1 LYSO St.Gobain + 1 LYSO Scionix crystals of
 1.5$\times$1.5$\times$13 cm$^3$ 
\item 1 LYSO St.Gobain crystal of 2$\times$2$\times$15 cm$^3$,
2 LYSO Scionix crystals of 2$\times$2$\times$13 cm$^3$.
\end{itemize}

The LFS from Zecotek is a Luthetium Fine Silicate crystal, 
with very similar properties to  LYSO.
Each crystal is wrapped with 100 $\mu$m of tyvek on the lateral faces,
leaving free both the front and end faces, thus allowing
to bring calibration light pulses through an external 
LED and a fast change of the photosensors readout.
Each APD is inserted in a PVC mask with the amplifier soldered on 
its anode and  then  mecchanically positioned inside
a stainless steel box closed by a PVC cap with only 
the electronic pins coming out for connection to
HV and readout cables. An external holder takes the  
PMs in position for the readout of the outer crystals 
while allowing to press the boxes  containing the APDs. 
The optical connection of the photosensors with the crystals 
is done with optical grease. The amplifiers are 
based on the MAR8A+ chip from Minicircuits, with a gain
factor of 25 and a bandwidth of 1 GHz.

\begin{figure}
\centering
\includegraphics[width=0.99\linewidth]{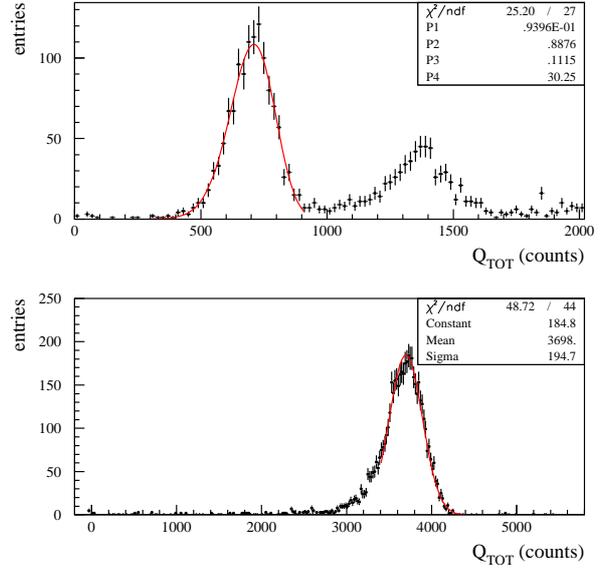}
\caption{Distribution of $Q_{TOT}$ for single electron
events at 100 MeV (top) with a logG fit superimposed
and at  500 MeV (bottom)  with a gaussian fit.}
\label{energy_response}
\end{figure}

\section{Test results with electron beams.}

We have taken data at the Beam Test Facility, BTF, of LNF for
two weeks in april 2009.
The matrix was positioned at the center of the beam axis with an area 
delimited by  a cross of two finger BC408 scintillators 
of $1 \times 0.5\times 5$ cm$^3$  dimensions, $f1,f2$.
In most of the tests, the fingers 
were aligned in such a way to define a beam spot of 
$1\times 1$ cm$^2$. In front of the fingers it was
also present a beam position monitor, BPM, of the BTF
group, consisting of  sixteen horizontal and  
vertical scintillator strips readout by two 
Multi Anode PMs. Each strip is built by three 1 mm diameter 
scintillating fibers thus providing an accuracy below 1 mm on the
beam localization.

We have triggered by using a replica of the  spill signal 
from the Linac adjustable from remote in order 
to correctly put the signals in time. Prototype signals 
were splitted and discriminated by means of the standard 
KLOE calorimeter SDS boards. We acquire data with the KLOE-2
daq system, VME based, reading out KLOE ADC 
and TDC boards with a sensitivity of 100 fC/count and 50 ps/count
respectively.

The BTF beam allows to set the average number of electrons, $<{\rm Ne}>$
arriving to the detector by adjusting dedicated
collimators. However, we could not lower
$<{\rm Ne}>$ below 1 for all beam energies and we cleaned 
the data offline by cutting on pulse height on the f1, f2 counters. 

\begin{figure}
\centering
\includegraphics[width=0.8\linewidth]{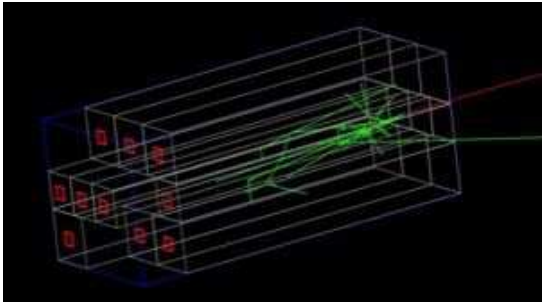}
\caption{Shower example generated by a single 
event of 100 MeV electron in the matrix as
shown by a full simulation based on Geant-4.}
\label{display_mc}
\end{figure}
Observing the response of the prototype to single electrons, 
we realized that the outer matrix was not properly working 
due to an unexpected 
optical cross-talk between crystals. This was discovered by 
pulsing a UV LED (at 380 nm) in the front face of each LYSO 
crystal to let it scintillate and testing, both at the scope  
and with a fast data taking, the amount of signal observed on 
the contiguous crystals. We observe large cross-talk only on 
the PbWO$_4$ of the outer matrix. We believe this to be due
by a cooperation between  a light leak through the tyvek 
and the different amplification gains between PMs and APDs. 
In the following,  we therefore report
only results related to the inner matrix.

\begin{figure}
\centering
\includegraphics[width=0.8\linewidth]{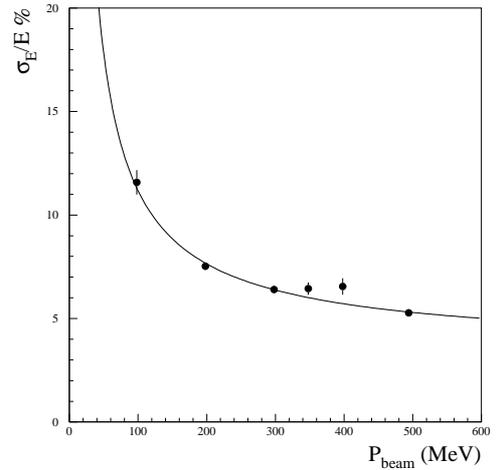}
\caption{Dependence of the energy resolution on beam
momentum}
\label{eres}
\end{figure}

By using the UV LED, we have first equalized each channel at 10\%
level by proper HV adjusting.
We have then calibrated the calorimeter response of each
channel with minimum ionizing particles, m.i.p., crossing
the calorimeter hortogonally to the crystal axis. To do this, 
we have collected cosmic ray runs with a dedicated  trigger defined by 
two plastic scintillators positioned one above and one below the
prototype. 
We get $\sigma_{ped}$ of 5 counts and a m.i.p. peak, $M_i$, 
of around 100 counts for the smaller size crystals. 
The statistical precision on the peak determination is $\sim$ 1 \%.
The total response of the detector is then defined as:
\begin{equation}
Q_{TOT} = \sum{ (Q_i-P_i)\times M_0/M_i},
\end{equation}
where $Q_i$ and $P_i$ are the collected charge and the pedestal
of the $i$-th channel, $M_0$ represents an average calibration
of all channels in counts and the calibration for the larger
crystals is corrected for the different size. In Fig.~\ref{energy_response},
we show the distribution of $Q_{tot}$ for a beam of 100
and 500 MeV respectively after having selected single
electron events with a cut on the finger scintillators. 
A surviving fraction
of events with more than 1 electron is still observed in
the matrix, expecially at low energies. 
We have fit the distribution corresponding
to one electron either with a simple gaussian, centered around
the peak, or with a logarithmic gaussian, logG, function
as follows:

\begin{equation}
{\rm N \cdot exp} (-\frac{1}{2\sigma_0^2} ln(1-\frac{\eta}{\sigma_E}(E-E_{peak}))^2
- \frac{\sigma_0^2}{2})
\end{equation}
where N is  a normalization factore, $\eta$ represents the asymmetry, 
$E_{peak}$ the most probable value of the  distribution,
$\sigma_0=\frac{2}{2.36}sinh^{-1}( 2.36\eta/2)$ and
 $\sigma_E = \frac{{\rm FWHM}}{2.36}$ is the resolution.

\begin{figure}
\centering
\includegraphics[width=0.9\linewidth]{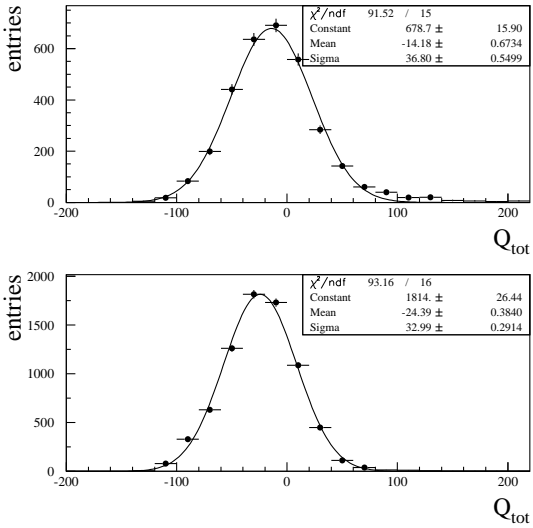}
\caption{Distribution of $Q_{TOT}$ for events without
electrons for runs with (up) 500 MeV, (down) 100 MeV  beam.}
\label{noise}
\end{figure}

By performing a linear fit to  the distribution of $E_{peak}$ vs $P_{beam}$,
we get a slope of 7.5$\pm$ 0.1 counts/MeV which sets the $M_i$ value
to be $\sim$ 16 MeV for a small crystal consistently  with an
expected energy loss of  $\sim$ 10 MeV/cm. At the running voltages of
410 V, the expected APD gain varies between 300-500 from which we
estimate the light yield to be between 500 and 800 p.e./MeV.

To understand the different terms of the energy dependence of
the energy resolution, we are carring out a full simulation
of the prototype based on Geant-4. In Fig.~\ref{display_mc}, we
report one example of 100 MeV electron shower as seen from
the MC program.  We have also simulated the 
dead space of 100 $\mu$m of tyvek between crystals 
and activated the optical transportation 
of photons. Studies
on photoelectron collection efficiency are still underways.
This simulation indicates that there is a large
leakage term between 5 and 4 \% from 100 to 500 MeV.

In Fig.~\ref{eres}, we show the energy dependence
of the energy resolution measured on data which has been
fit with the following equation:
\begin{equation}
\sigma_E/E = a \oplus {\rm b/(E/GeV)} \oplus \rm{c}/\sqrt{\rm E/GeV},
\end{equation}
where, accordingly to MC, we have fixed the constant term to be 5 \%. 
We found $b=1.1\%$ and $c=1.4\%$ when using the gaussian fits to
the spectra. If we repeat this procedure, for the fits with the logG
function, we get $b=0.8$\% and $c=2.4$\%.

We have investigated the large b/E term by measuring the total
detector noise with a gaussian fit to $Q_{TOT}$ in events
without any electron beam impinging (see Fig.~\ref{noise}). 
We find $\sigma_{Q}=$ 4.2$\div$4.8 MeV which is slightly larger than 
the incoherent sum of $\sigma_{ped}$ resulting to be 3.6$\div$3.8 MeV.
A not negligible coherent noise is present and a much 
smaller (1/2) noise level has been previously  
measured in the electronic laboratory. 
However, the noise does not fully explain the large $b$ term found.
We are still investigating the origin of this contribution.

We have then determined the position resolution by comparing
the reconstructed centroid done with the crystals with 
the position provided by the BPM of BTF. 
The centroids were defined as $X_{pos} = \sum{ Q_i X_i}/Q_{tot}$.
We observe a position resolution of 2.8 $\div$ 3 mm at 500 MeV,
which is slightly better than the crystal pitch as expected 
by the application of the centroid method.

\begin{figure}
\centering
\includegraphics[width=0.8\linewidth]{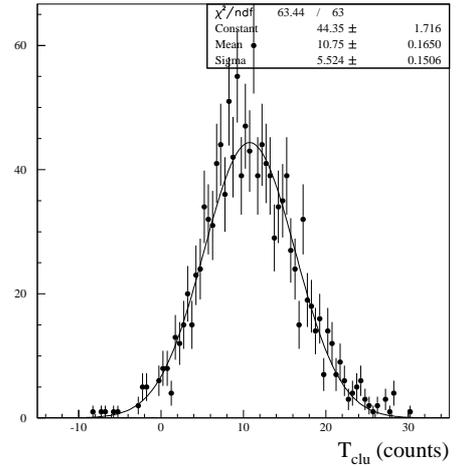}
\caption{Distribution of the average timing reconstructed
with the inner matrix for 100 MeV electron beam.}
\label{timing}
\end{figure}

We have finally reconstructed  the calorimeter timing after correcting it, 
event by event, for the arrival time of the electrons in the LINAC spill.
This was done by measuring the timing with the scintillators f1,f2.
A very small pulse-height dependence on the calorimeter timing
remains due to the usage of Constant Fraction Discriminators of KLOE.
The weighted energy average over all calorimeter, $T_{clu}$,  was done
after subtracting the average T$_0$ of each cell. 
In Fig.~\ref{timing}, the distribution of $T_{clu}$ is shown for
100 MeV electron beam. A clean gaussian response is observed with
a time resolution, $\sigma_T$, of  $\sim$ 265 ps at 100 MeV. 
A similar study at other energies provided compatible results 
(e.g. $\sigma_T \sim 245$ ps at 500 MeV). We have also measured the
time jitter of the f1,f2 scintillators, by studying their time
difference, which resulted to be of $\sim$ 240 ps thus explaining
the week energy dependence of $\sigma_T$ and pushing for a new
measurement with a much reduced time-jitter.

\section*{Acknowledgments}
The authors are indebt to many people for the succesfull realization
of the matrix. In particular, we thank M.~Lobello from Roma-3 university for
the mechanical drawings, all the LNF mechanical shop for the
realization of the support and APD boxes, expecially G.~Bisogni, 
U.~Martini and A.~De Paolis. We also thank the BTF staff for
providing the beam time and  
B.~Buonomo for great support during the test beam data taking.
The realization of the preamplifiers was done in collaboration
with E.Reali from  Roma-2 university.



\begin{thebibliography}{10}
\expandafter\ifx\csname url\endcsname\relax
  \def\url#1{\texttt{#1}}\fi
\expandafter\ifx\csname urlprefix\endcsname\relax\def\urlprefix{URL }\fi

\bibitem{Dafne}
P.Raimondi, in: ``Crab Waist Collisions in DA$\Phi$NE AND SUPER-B DESIGN'',
Proceedings of EPAC08, Genoa, Italy (2008).

\bibitem{kloe2_loi}
F.~Bossi et al, for the KLOE-2 collaboration,''A proposal for the Roll-In
of the KLOE-2 detector'', LNF-Internal Note, 07/19 (2007). 

\bibitem{kloe_cimento}
F.~Bossi et al, for the KLOE collaboration, ``Precision Kaon and Hadron
Physics with KLOE'', Riv. Nuovo Cimento 031:531-623 (2009)

\bibitem{ksgg}
The KLOE collaboration, ``Measurement of the BR($K_S \to \gamma \gamma$)
using a pure $K_S$ beam with the KLOE detector'', JHEP 0805:051 (2008).
\end{thebibliography}
\end{document}